\documentclass[12pt]{article}
\begin{document}
\title{Mass Generation and Non Commutative Spacetime}
\author{B.G. Sidharth\\
International Institute for Applicable Mathematics \& Information Sciences\\
Hyderabad (India) \& Udine (Italy)\\
B.M. Birla Science Centre, Adarsh Nagar, Hyderabad - 500 063
(India)}
\date{}
\maketitle
\begin{abstract}
In this paper we show how it is possible to obtain mass generation
in the context of non Abelian gauge field theory, using a non
commutative spacetime. This is further confirmed by the modified
dispersion relation that results from such a geometry.
\end{abstract}
\section{Introduction}
It is well known that Hermann Weyl's original phase transformation
proposal was generalized, so that the global or constant phase of
$\lambda$ was considered to be a function of the coordinates
\cite{uof,moriyasu,jacob,greiner}.\\
As is well known this leads to a covariant gauge derivative. For
example, the transformation arising from $(x^\mu) \to (x^\mu +
dx^\mu)$,
\begin{equation}
\psi \to \psi e^{-\imath \lambda}\label{Ee1}
\end{equation}
leads to the familiar \index{electromagnetic}electromagnetic
potential gauge,
\begin{equation}
A_\mu \to A_\mu - \partial_\mu \lambda\label{Ee2}
\end{equation}
The above transformation, ofcourse, is a \index{symmetry}symmetry
transformation. In the transition from (\ref{Ee1}) to (\ref{Ee2}),
we expand the exponential,
retaining terms only to the first order in coordinate differentials.\\
Let us now consider the gauge field in some detail. As is known this
could be obtained as a generalization of the above phase function
$\lambda$ to include fields with internal degrees of freedom. For
example $\lambda$ could be replaced by $A_\mu$ given by
\cite{moriyasu}
\begin{equation}
A_\mu = \sum_{\imath} A^\imath_\mu (x)L_\imath ,\label{Eex1}
\end{equation}
The \index{gauge field}gauge field itself would be obtained by using
Stoke's Theorem and (\ref{Eex1}). This is a very well known
procedure: considering a circuit, which for simplicity we can take
to be a parallelogram of side $dx$ and $dy$ in two dimensions, we
can easily deduce the equation for the field, viz.,
\begin{equation}
F_{\mu \nu} = \partial_\mu A_\nu - \partial_\nu A_\mu - \imath q
[A_\mu , A_\nu ],\label{Eex2}
\end{equation}
$q$ being the \index{gauge field}gauge field coupling constant.\\
In (\ref{Eex2}), the second term on the right side is typical of a
non Abelian \index{gauge field}gauge field. In the case of the U(1)
\index{electromagnetic}electromagnetic field, this latter term vanishes.\\
Further as is well known, in a typical Lagrangian like
\begin{equation}
\mathit{L} = \imath \bar \psi \gamma^\mu D_\mu \psi - \frac{1}{4}
F^{\mu \nu} F_{\mu \nu} - m \bar \psi \psi\label{Eex3}
\end{equation}
$D$ denoting the Gauge \index{covariant derivative}covariant
derivative, there is no \index{mass}mass term for the field
\index{Boson}Bosons. Such a \index{mass}mass term in (\ref{Eex3})
must have the form $m^2 A^\mu A_\mu$ which unfortunately is not
Gauge invariant.\\
This was the shortcoming of the original
\index{Yang-Mills}Yang-Mills Gauge Theory: The Gauge Bosons would be
\index{mass}massless and hence the need for a \index{symmetry
breaking}symmetry breaking,
\index{mass}mass generating mechanism.\\
The well known remedy for the above situation has been to consider,
in analogy with \index{superconductivity}superconductivity theory,
an extra phase of a self coherent system (Cf.ref.\cite{moriyasu} for
a simple and elegant treatment and also refs. \cite{jacob} and
\cite{taylor}). Thus instead of the \index{gauge field}gauge field
$A_\mu$, we consider a new phase adjusted \index{gauge field}gauge
field after the \index{symmetry}symmetry is broken
\begin{equation}
W_\mu = A_\mu - \frac{1}{q} \partial_\mu \phi\label{Eex4}
\end{equation}
The field $W_\mu$ now generates the \index{mass}mass in a self
consistent manner via a Higgs mechanism. Infact the kinetic energy
term
\begin{equation}
\frac{1}{2} |D_\mu \phi |^2\quad ,\label{Eex5}
\end{equation}
where $D_\mu$ in (\ref{Eex5}) denotes the Gauge \index{covariant
derivative}, now becomes
\begin{equation}
|D_\mu \phi_0 |^2 = q^2|W_\mu |^2 |\phi_0 |^2 \, ,\label{Eex6}
\end{equation}
Equation (\ref{Eex6}) gives the \index{mass}mass in terms of the ground state $\phi_0$.\\
The whole point is as follows: The \index{symmetry breaking}symmetry
breaking of the \index{gauge field}gauge field manifests itself only
at short length scales signifying the fact that the field is
mediated by particles with large \index{mass}mass. Further the
internal \index{symmetry}symmetry space of the \index{gauge
field}gauge field is broken by an external constraint: the wave
function has an intrinsic relative phase factor which is a different
function of spacetime coordinates compared to the phase change
necessitated by the minimum coupling requirement for a free particle
with the gauge potential. This cannot be achieved for an ordinary
point like particle, but a new type of a physical system, like the
self coherent system of \index{superconductivity}superconductivity
theory now interacts with the \index{gauge field}gauge field. The
second or extra term in (\ref{Eex4}) is effectively an external
field, though (\ref{Eex6}) manifests itself only in a relatively
small spatial interval. The $\phi$ of the Higgs field in
(\ref{Eex4}), in analogy with the phase function of  \index{Cooper
pairs}Cooper pairs of \index{superconductivity}superconductivity
theory comes with a
\index{Landau-Ginzburg}Landau-Ginzburg potential $V(\phi)$.\\
Let us now consider in the \index{gauge field}gauge field
transformation, an additional phase term, $f(x)$, this being a
scalar. In the usual theory such a term can always be gauged away in
the \index{U(1)}U(1) \index{electromagnetic}electromagnetic group.
However we now consider the new situation of a
\index{noncommutative}noncommutative geometry viz.,
\begin{equation}
\left[dx^\mu , dx^\nu \right] = \Theta^{\mu \nu} \beta , \beta \sim
0 (l^2)\label{Eex7}
\end{equation}
where $l$ denotes a minimum \index{spacetime}spacetime cut off.
Equation (\ref{Eex7}) is infact \index{Lorentz}Lorentz covariant.
Then the $f$ phase factor gives a contribution to the second order
in coordinate differentials,
$$\frac{1}{2} \left[\partial_\mu B_\nu - \partial_\nu B_\mu \right] \left[dx^\mu , dx^\nu \right]$$
\begin{equation}
+ \frac{1}{2} \left[\partial_\mu B_\nu + \partial_\nu B_\mu \right]
\left[dx^\mu dx^\nu + dx^\nu dx^\mu \right]\label{Eex8}
\end{equation}
where $B_\mu \equiv \partial_\mu f$.\\
As can be seen from (\ref{Eex8}) and (\ref{Eex7}), the new
contribution is in the term which contains the commutator of the
coordinate differentials, and not in the symmetric second term.
Effectively, remembering that $B_\mu$ arises from the scalar phase
factor, and not from the non-Abelian \index{gauge field}gauge field,
in equation (\ref{Eex2}) $A_\mu$ is replaced by
\begin{equation}
A_\mu \to A_\mu + B_\mu = A_\mu + \partial_\mu f\label{Eex9}
\end{equation}
Comparing (\ref{Eex9}) with (\ref{Eex4}) we can immediately see that
the effect of noncommutativity is precisely that of providing a new
\index{symmetry breaking}symmetry breaking term to the \index{gauge
field}gauge field, instead of the $\phi$ term, (Cf.refs.
\cite{cr39,ijmpe}) a term not belonging to the
\index{gauge field}gauge field itself.\\
On the other hand if we neglect in (\ref{Eex7}) terms $\sim l^2$,
then there is no extra contribution coming from (\ref{Eex8}) or
(\ref{Eex9}), so that we are in the usual non-Abelian \index{gauge
field}gauge field theory, requiring a broken
\index{symmetry}symmetry to obtain an equation like (\ref{Eex9}).\\
To see this in greater detail, we note that, as shown by the author
in early 2000 \cite{bgsdiscrete} given a minimum length $l$, the
energy momentum relation gets modified. The usual Quantum Mechanical
commutation relations get modified and now become
\begin{equation}
[x,p] = \hbar' = \hbar [1 + \left(\frac{l}{\hbar}\right)^2 p^2]\,
etc\label{5He2}
\end{equation}
(Cf. also ref.\cite{bgsust}). (\ref{5He2}) shows that effectively
$\hbar$ is replaced by $\hbar'$. So, in units, $\hbar = 1 = c$,
$$E = [m^2 + p^2 (1 + l^2 p^2)^{-2}]^{\frac{1}{2}}$$
or, the energy-momentum relation leading to the Klein-Gordon
Hamiltonian is given by,
\begin{equation}
E^2 = m^2 + p^2 - 2l^2 p^4,\label{5He3}
\end{equation}
neglecting higher order terms. This is the so called Snyder-Sidharth
Hamiltonian for Bosons \cite{glinka}. (It may be mentioned that some
other authors have since ad hoc taken a third power of $p$, and so
on \cite{myers}. However we should remember that these were mostly
phenomenological approaches.)\\
For Fermions the analysis can be more detailed, in terms of Wilson
lattices \cite{mont}. The free Hamiltonian now describes a
collection of harmonic fermionic oscillators in momentum space.
Assuming periodic boundary conditions in all three directions of a
cube of dimension $L^3$, the allowed momentum components are
\begin{equation}
{\bf q} \equiv \left\{q_k = \frac{2\pi}{L}v_k; k = 1,2,3 \right\},
\quad 0 \leq v_k \leq L - 1\label{4.59}
\end{equation}
(\ref{4.59}) finally leads to
\begin{equation}
E_{\bf q} = \pm \left(m^2 + \sum^{3}_{k=1} a^{-2} sin^2
q_k\right)^{1/2}\label{4.62}
\end{equation}
where $a = l$ is the length of the lattice, this being the desired
result leading to
\begin{equation}
E^2 = p^2e^2+m^2c^4 + \alpha l^2p^4\label{ex} \end{equation}
(\ref{ex}) shows that $\alpha$ is positive, that is for Fermions the
Snyder-Sidharth Hamiltonian is given by (\ref{ex}).
\section{A Modified Dirac Equation}
Once we consider a discrete spacetime structure, the energy momentum
relation, as noted, gets modified \cite{ijmpe2,mont} and we have in
units $c = 1 = \hbar$,
\begin{equation}
E^2 - p^2 - m^2 - l^2 p^4 = 0\label{6ce1}
\end{equation}
$l$ being the Planck length. Let us now consider the Dirac equation
\begin{equation}
\left\{ \gamma^\mu p_\mu - m\right\} \psi \equiv \left\{\gamma^\circ
p^\circ + \Gamma \right\} \psi = 0\label{6ce2}
\end{equation}
If we include the extra effect shown in (\ref{6ce1}) we get
\begin{equation}
\left(\gamma^\circ p^\circ + \Gamma + \beta l p^2\right) \psi =
0\label{6ce3}
\end{equation}
$\beta$ being a suitable matrix.\\
Multiplying (\ref{6ce3}) by the operator
$$\left(\gamma^\circ p^\circ - \Gamma - \beta l p^2\right)$$
on the left we get
\begin{equation}
p^2_0 - \left(\Gamma \Gamma + \left\{\Gamma \beta + \beta
\Gamma\right\} + \beta^2 l^2 p^4\right\} \psi = 0\label{6ce4}
\end{equation}
If (\ref{6ce4}), as in the usual theory, has to represent
(\ref{6ce1}), then we require that the matrix $\beta$ satisfy
\begin{equation}
\Gamma \beta + \beta \Gamma = 0, \quad \beta^2 = 1\label{6ce5}
\end{equation}
It follows that,
\begin{equation}
\beta = \gamma^5\label{6ce6}
\end{equation}
Using (\ref{6ce6}) in (\ref{6ce3}), the modified Dirac equation the
so called Dirac-Sidharth equation \cite{glinka} finally becomes
\begin{equation}
\left\{\gamma^\circ p^\circ + \Gamma + \gamma^5 l p^2\right\} \psi =
0\label{6ce7}
\end{equation}
Owing to the fact that we have \cite{bd2}
\begin{equation}
P \gamma^5 = -\gamma^5 P\label{6ce8}
\end{equation}
It follows that the modified Dirac equation (\ref{6ce7}) is not
invariant under reflections.\\
We can also see that due to the modified Dirac equation
(\ref{6ce7}), there is no additional effect on the anomalous
gyromagnetic ratio. This is because, in the usual equation from
which the magnetic moment is determined \cite{merz} viz.,
$$\frac{d\vec{S}}{dt} = -\frac{e}{\mu c} \vec{B} \times \vec{S},$$
where $\vec{S} = \hbar \sum/2$ is the electron spin operator, there
is now an extra term
\begin{equation}
\left[\gamma^5, \sum\right]\label{6ce9}
\end{equation}
However the expression (\ref{6ce9}) vanishes by the property of the
Dirac matrices.\\
We point out that using the modified dispersion relation (\ref{ex}),
for a massless particle, $m = 0$, and identifying the extra term
$l^2p^4$ as being due to a mass $\delta m$, we can easily deduce
that, restoring proper units,
$$\frac{c^2}{\hbar^2} l^2 p^4 = \Delta E^2 = \delta m^2 c^4,$$
Whence,
\begin{equation}
\delta m = \frac{\hbar}{cl} \quad \mbox{or} \quad l =
\frac{\hbar}{c\delta m}\label{e26} \end{equation} This shows that
$l$ is the Compton wavelength for this mass $\delta m$ or
alternatively if $l$ is the Compton wavelength, then we deduce the
mass, now generated from the extra effect. This is another
demonstration of mass generation from $O(l^2)$ effects as seen in
Section 1, without requiring a Higgs mechanism. If, for example, $l$
were the Planck length, then $\delta m$ would be the Planck mass
(and vice versa).
\section{Remarks}
Finally, we would like to make two remarks. Glinka has used the
Snyder-Sidharth Hamiltonian (\ref{5He3}), to show that a consistent
structure is \cite{glinka2} ,
\begin{equation}\label{eq-remark1}E = \frac{p^2}{2m} + m \quad \quad (c = 1)\end{equation}
This semi-relativistic solution too, brings out the extra mass (or
energy) term.\\
There is another way of looking at all this. Because of Quantum Theory, the relativistic formula,
$$(\Delta x)^2 -(\Delta t)^2 = 0$$
goes over to, as is well known,
\begin{equation}\label{eq-remark2}(\Delta x)^2 -(\Delta t)^2 \leq \frac{1}{m^2} \quad \quad (\hbar = 1)\end{equation}
Effectively, the velocity of light $c \,(=1)$ becomes $c'$, given by from (\ref{eq-remark2}),
\begin{equation}\label{eq-remark3} {c'}^{2} \leq\, 1 + \frac{1}{m^2 \Delta^2}= 2 \end{equation}
as $\Delta t \geq \frac{1}{m},$ the Compton time.\\So, the usual energy $E\,=\, m c^2$ becomes $E'$  where 
\begin{equation}\label{eq-remark4} E\;< E'\; \leq \; 2E.\end{equation}
There is thus an increase in energy in the extreme relativistic case. This conclusion (\ref{eq-remark4}) can also be deduced from (\ref{eq-remark1}) if we use (\ref{eq-remark3}). Then we get back (\ref{eq-remark4}).\\
Secondly, if we started out with a massless particle like a photon,
then using the fact that $E = p = m (c = 1)$, we get directly from
(\ref{5He3}),
$$l \sim \frac{1}{m}$$
This is in agreement with earlier results \cite{tduniv}, giving the
photon a mass of $\sim 10^{-65}gm$ and $l \sim$ radius of the
universe $\sim 10^{28}cm$.


\begin{thebibliography}{99}
\bibitem {uof} Sidharth, B.G. (2005). \emph{The Universe of Fluctuations} (Springer,
Netherlands).
\bibitem {moriyasu}  Moriyasu, K. (1983). \emph{An Elementary Primer for Gauge Theory}
(World Scientific, Singapore).
\bibitem {jacob} Jacob, M. (1974). \emph{Physics Reports, Reprint Volume} (North-Holland,
Amsterdam).
\bibitem {greiner} Greiner, W. and Reinhardt, I. (1995).
\emph{Gauge Theory of Weak Interactions} (Springer-Verlag, Berlin).
\bibitem {taylor} Taylor, J.C. (1978). \emph{Gauge Theories of Weak Interactions} (Cambridge
University Press, Cambridge).
\bibitem {cr39}  Sidharth, B.G. (2004). \emph{Proceedings of the Fifth International Symposium} on
\emph{``Frontiers of Fundamental Physics''} (Universities Press,
Hyderabad).
\bibitem {ijmpe} Sidharth, B.G. (2006). \emph{Int.J.SCI.} January 2006, pp.42ff.
\bibitem {bgsdiscrete} Sidharth, B.G. (2000). \emph{Chaos, Solitons and Fractals} 11, 2000.
\bibitem {bgsust} Sidharth, B.G. (2005). \emph{Chaos, Solitons and Fractals} 25, pp.965--968.
\bibitem {glinka} Glinka, L. (2008). \emph{arXiv:hep-th
0812.0551v1}.
\bibitem {myers} Myers, R.C. and Pospalov, M. (2003).
\emph{Phys.Rev.Lett.} 90:211601.
\bibitem {mont} Montvay, I. and Munster, G. (1994). \emph{Quantum Fields on a
Lattice} (Cambridge University Press) pp.174ff.
\bibitem {ijmpe2} Sidharth, B.G. (2005). \emph{Int.J.Mod.Phys.E.} 14, (6), pp.923ff.
\bibitem {bd2} Bjorken, J.D. and Drell, S.D. (1965). \emph{Relativistic Quantum Fields}
(McGraw-Hill Inc., New York), pp.44ff.
\bibitem {merz} E. Merzbacher. (1970). \emph{Quantum Mechanics} (Wiley, New
York); Greiner, W. (1983). \emph{Relativistic Quantum Mechanics:
Wave Equation} 2nd Ed. (Springer);  Greiner, W. and Reinhardt, J.
(1987). \emph{Quantum Electrodynamics} 3rd Ed. (Springer).
\bibitem {glinka2} Glinka, L.A. (2009). \emph{0811.0551} and \emph{Apeiron} 2, April
2009.
\bibitem {tduniv} Sidharth, B.G. (2008) \emph{The Thermodynamic
Universe} (World Scientific), Singapore.
\end{thebibliography}
\end{document}